\documentclass{nature_with_graphic}

\usepackage{graphicx} 
\usepackage{float} 
\usepackage{subfigure}
\usepackage{caption}
\usepackage{textcomp}
\usepackage{upgreek}
\usepackage{amsmath}
\usepackage{mathabx}
\usepackage{braket}
\usepackage[dvipsnames]{xcolor}
\title{Symmetry mediated tunable molecular magnetism on a 2D material}
\author{Yuqi Wang$^{1,2\ast}$, Soroush Arabi$^{1,2,3}$,
Klaus Kern$^{1,4}$ \& Markus Ternes$^{2,3\ast}$}

\begin{document}
%\linenumbers

\maketitle

\begin{affiliations}
 \item Max Planck Institute for Solid State Research, Heisenbergstraße 1, D-70569 Stuttgart, Germany.
 \item Peter-Gr\"{u}nberg-Institute, Forschungszentrum J\"{u}lich, D-52425 J\"{u}lich, Germany.
 \item Institute of Physics, RWTH Aachen University, D-52074 Aachen, Germany.
 \item Institut de Physique, \'{E}cole Polytechnique F\'{e}d\'{e}rale de Lausanne, CH-1015 Lausanne, Switzerland.

\noindent $^\ast$corresponding authors:  yuqi.wang@fkf.mpg.de; ternes@physik.rwth-aachen.de

\end{affiliations}

%approx 200 words!
\begin{abstract}
The induction of unconventional superconductivity by twisting two layers of graphene a small angle was groundbreaking\cite{cao2018unconventional}, and since then has attracted widespread attention to novel phenomena caused by lattice or angle mismatch between two-dimensional (2D) materials\cite{cao2018correlated}.
While many studies address the influence of angle mismatch between layered 2D 
materials\cite{sharpe2019ferromagnetism_in_twisted_bilayer_graphene,
chen2019Mott_insulator_trilayer_graphene,
chen2019gate_tunable_superconductivity_in_trilayer_graphene_hBN_moire_superlatti
ces}, the impact of the absorption alignment on the physical properties of 
planar molecules on 2D substrates has not been studied in detail.
Using scanning probe microscopy (SPM) we show that individual cobalt 
phthalocyanine (CoPc) molecules adsorbed on the layered superconductor 
2H-NbSe$_2$ change drastically their charge and spin state when the symmetry 
axes of the molecule and the substrate are twisted with respect to each other.
The CoPc changes from an effective spin-1/2 as found in gas-phase\cite{mugarza2012DFT_MPC_gasphase_onAg(100)} to a molecule with non-magnetic ground-state. 
On the latter we observe a singlet-triplet transition originating from an antiferromagnetic interaction between the central-ion spin and a distributed magnetic moment on the molecular ligands. 
Because the Ising superconductor 2H-NbSe$_2$ lacks inversion symmetry and  has large spin-orbit coupling\cite{xi2016NbSe2isingSC} this intramolecular magnetic exchange has significant non-collinear Dzyaloshinskii–Moriya (DM)\cite{dzialoshinskii1957thermodynamic,moriya1960anisotropic} contribution.
\end{abstract}

Symmetry as a fundamental concept enables to classify properties of molecules and materials such as their optical activity, electronic bandstructure, or vibration modes\cite{cotton2003chemical}. 
%optical activity--chiral molecules and rotate plane-polarized light;
The molecular symmetry can be changed without modifying its structure by adsorbing the molecule on a sample with a dissimilar point-group.
%
%For a single molecule, to tune molecular symmetry without changing its original structure, one can use substrates with dissimilar point-group. 
In particular, low dimensional 2D materials are potentially interesting as platforms because of their intrinsic slowly decaying long-range interactions and the resulting extended coherence\cite{menard2015coherent_length_YSR_Co_NbSe2}.  
Especially the surface of H phase transition metal dichalcogenides (TMDs) provide broken in-plane inversion symmetry and significant spin-orbit coupling induced by the central 4\textit{d} metal ions.
%\cite{patera2019Nature_AFM_CuPc_NaCl}
%\cite{altenburg2015FePc_graphene}
%2D materials and in particular transition metal dichalcogenides (TMDs) are potentially interesting as platforms for the study of long range magnetic interactions, due to their slowly decaying in-plane coupling strength and the resulting extended coherence\cite{menard2015coherent_length_YSR_Co_NbSe2}. 
%Furthermore the central 4\textit{d} transition metal ions induce significant spin-orbit coupling in TMDs. 
These properties together are key for introducing non-collinear magnetism via antisymmetric DM exchange interactions in adsorbed magnetic systems\cite{bode2007chiral}.

We choose CoPc, a highly symmetric metal-organic complex with flat adsorption geometry, on superconducting 2H-NbSe$_2$ to explore the influence of symmetry on the magnetic properties of metal-organic molecules by means of SPM.
We find two stable adsorption sites of the molecules (Fig.~1a, c) which differ in their in-plane orientation: 
The molecule is either aligned or twisted by $15^{\circ}$ with respect to the main surface directions (Fig.~1d, f).
Constant-height SPM images also show slight differences in the topographic appearance of the two types hinting to a distinction between their electronic structures. 
At first glance both types of CoPc molecules have retained their cross-like appearance, however, a closer inspection reveals that the topography of the molecules has only mirror symmetry. 

We can rationalize the observation of the two differently adsorbed CoPc molecules by noticing that the difference between $C_{3v}$ symmetry of the surface and $C_{4v}$ symmetry %(4-fold rotation as well as two $\sigma_v$ and two $\sigma_d$ mirror planes) 
of the molecules completely breaks all nontrivial rotational symmetries letting the mirror symmetries the only retaining ones of the system (for details see SOM).
Therefore, to reach maximum symmetry, one of the three $\sigma_v$ mirror planes of the sample can be either aligned to one of the two $\sigma_v$ or the two $\sigma_d$ mirror planes of the CoPc molecules, resulting naturally in the two different adsorption geometries with $15^{\circ}$ rotational difference.
Those molecules in which the $\sigma_v$ of the sample is aligned with their $\sigma_d$ have magnetic properties similar to CoPc in the gas phase with an effective spin $S=1/2$ that originates from an unpaired electron at the central Co$^{2+}$ ion\cite{kezilebieke2018coupledCoPcNbSe2}. 
We label these molecules CoPc$_{\rm d}$. 
%At temperatures below the characteristic Kondo temperature $T_K$ magnetic exchange interactions between the doublet state and the conduction electrons of the 2D sample should lead to the emergence of a Kondo resonance in which the conduction electrons screen the local magnetic moment of the molecule forming a many electron singlet state%\cite{kondo1968effect,madhavan1998KondoResonanceSTM,ternes2015spinSTM,ternes2017SpinSTM}.
%\cite{kondo1968effect,ternes2017SpinSTM}. 
%However, because 2H-NbSe$_2$ is a superconductor (SC) 
%the situation becomes more complex:
%When the Kondo screening energy $k_BT_K$, with $k_B$ as  the Boltzmann constant, is much smaller than the Cooper pair binding energy of the sample $\Delta_S$, i.\,e.\ $k_BT_K\ll\Delta_S$, the opening of the SC gap hinders Kondo screening by depleting the available electronic states at the Fermi energy\cite{franke2011competition_YSR_Kondo}.
%Scattering of Cooper pairs at this now unscreened magnetic moment leads to a pair of Yu-Shiba-Rusinov (YSR)  states at an energy $\pm E_{\rm YSR}$ within the SC gap\cite{yu1965_Y_YSR,shiba1968_S_YSR,rusinov1969_R_YSR,yazdani1997YSR_STM,heinrich2018YSRreview} (Fig.~1k).
%
Contrarily, molecules in which the $\sigma_v$ of the sample and of the molecule are aligned couple stronger to the substrate and enable charge transfer between the molecular orbitals and the sample.
%The intra-molecular antiferromagnetic interaction $J_{ST}$ between the central spin on the Co and the distributed spin on the ligand (Fig.~1l) results in a molecular singlet ground state and a singlet-triplet transition which manifest itself as steps in differential conductance ($dI/dV$)  measurements at $|eV|>J_{ST}$\cite{kondo1968effect,mishra2020_Singlet_Triplet_transition_nanographene} (Fig.~1m).
We label these molecules as  CoPc$_{\rm v}$.

We now characterize in detail the magnetic state of CoPc$_{\rm d}$ by $dI/dV$ spectroscopy using a Pb-coated SC tip with an effective gap of $\Delta_T\approx1.15$~meV (for details see methods).   
Placing the tip of our SPM over the bare sample we observe a gap of $\pm(\Delta_S+\Delta_T)$ due to SC -- SC tunneling between tip and sample ($\Delta_S\approx 1.3$~meV) while we measure a pair of peaks at $\approx \pm 1.8$~meV on the molecule (Fig.~2a,~b). 
These peaks originate from the scattering of Cooper pairs at the unscreened magnetic moment of the CoPc$_{\rm d}$ molecule leading to a pair of Yu-Shiba-Rusinov (YSR) states within the SC gap of the surface\cite{yu1965_Y_YSR,shiba1968_S_YSR,rusinov1969_R_YSR,heinrich2018YSRreview}.
We find good agreement with the measured data when simulating these YSR states using a scattering model in which the impurity is treated in classical approximation with an effective magnetic moment of $\frac{1}{2}\pi\rho_SJ_SS=-0.60 \pm 0.02$ and where $J_S$ is the coupling strength between the Co$^{2+}$ ion and the sample, $\rho_S$ is the density of electron states of the sample in the normal conducting phase, and $S$ is the effective spin of the central ion (for details see SOM)\cite{salkola2002YSR_fit}. 
The asymmetric intensity between the peaks at positive and negative bias indicates that particle-hole symmetry is broken which we account for by an additional Coulomb scattering of $\pi\rho_SU=0.28 \pm 0.02$. 

In order to infer the spin of the CoPc$_{\rm d}$, we apply a magnetic field $B\geq 5$~T perpendicular to the sample surface that is strong enough to suppress SC in tip and sample.
Contrarily to the $B=0$ data, we now observe split peaks around zero bias (Fig.~2c), typical for an $S=1/2$ spin in the weak coupling Kondo regime where the Zeeman energy $E_Z = g\mu_B B$ is larger than the Kondo energy $k_BT_K$ ($\mu_B$ is the Bohr magneton and $k_B$ is the Boltzmann constant)\cite{Zhang13_split_Kondo}.   
The Kondo temperature $T_K$ is the characteristic temperature below which magnetic exchange interactions between the doublet state and the conduction electrons of the sample screen the local magnetic moment of the molecule forming a many-electron singlet state\cite{kondo1968effect,ternes2017SpinSTM}. 
A linear regression of the peak splitting leads to a Land\'{e} $g$-factor of $1.54 \pm 0.02$, significantly smaller than the one for a free electron (Fig.~2d). 
The interception of the fit with the abscissa is not at the origin but at a $B_K=0.67\pm0.19$~T.
In linear approximation, $B_K$ is the minimal field strength necessary for splitting the Kondo singlet state and enables the estimation of $T_K\approx g\mu_B B_K/k_B=0.77\pm0.24$~K\cite{vzitko2009splitting_Kondo}. 
Because the Kondo screening energy is much smaller than the Cooper pair binding energy of the sample, i.\,e.\
$\Delta_S\gg k_BT_K$, the opening of the SC gap at $B=0$ hinders Kondo screening by depleting the available electronic states at the Fermi energy, in perfect agreement with the appearance of the YSR-states\cite{franke2011competition_YSR_Kondo}.

We now turn  our interest towards the CoPc$_{\rm v}$ molecules. 
%We continue now to measure $dI/dV$ spectra on the CoPc(ST) molecules. 
On these molecules we detect strong spectroscopic features at $|V|\approx23-25$~mV, but %we detect %at lower absolute energies
neither a Kondo peak close to zero bias nor YSR states inside the SC gap (Fig.~3a, b). 
Indeed, comparing the SC gap measured on the bare surface and on the molecule reveal no detectable difference at $B=0$ (Fig.~3b). 
In contrast to the observation on CoPc$_{\rm d}$, even at $B$-fields large enough to suppress SC, we observe only a flat and featureless spectrum suggesting that CoPc$_{\rm v}$ is not $S=1/2$, but has a non-magnetic ground-state (Fig.~3b). 
However, our observation of a strong conductance increase at higher absolute biases is a clear indicator for inelastic excitations\cite{ternes2017SpinSTM}.
At $B=0$ the $dI/dV$ spectrum shows sharp peaks on top of the steps which are induced by the convolution with the SC tip and sample spectrum\cite{heinrich2013Spin_excitation_SC_tip}.
This convolution also leads to an apparent shift of the excitation energy by $\Delta_{T}+\Delta_{S}$.
%and disappear at $B\geq 5$~T where SC is quenched. 
By increasing the $B$-field we observe that the $dI/dV$ steps or, equivalently, the peaks in $d^2I/dV^2$ successively split (Fig.~3c--f), proving that the feature is of magnetic origin. 
Note, our observations are not compatible with vibrational excitations which have been observed on CoPc molecules adsorbed on Ag(110) at similar energy but with much lower intensity\cite{chiang2014IETS_COtip}.

Kelvin-Probe measurements on top of the Co$^{2+}$ ion show only a negligible change of the local work function, but a stronger hybridization between the CoPc$_{\rm v}$ molecule and sample (see SOM).
This suggests that the change of symmetries by the slight change of orientation is accompanied by a charge transfer between the ligands of the CoPc$_{\rm v}$ molecule and the substrate. 
This transfer induces an additional magnetic moment which interacts antiferromagnetically with the moment at the central metal ion leading to the observed singlet ground state of CoPc$_{\rm v}$. 
However, Heisenberg-like interactions between the two spins alone would lead to a triplet of excitations at high enough $B$-fields (Fig.~3g). 
In contrast, we observe a splitting in only two distinguishable excitations. 
Remarkably, the excitations  at lower absolute energy have about twice the intensity of the ones at higher absolute energy.
This points to additional non-collinear interactions between both spins. To get a deeper understanding we model the excitation energy using the following Hamiltonian:
\begin{equation}
\hat H _{\rm CoPc_{\rm v}} = \sum\limits_{i=1,2}g\mu_B\hat S_{z}^{i}{B} + J_{ST}\cdot \hat{\bf S}_{1} \cdot \hat{\bf S}_{2} + \vec{D}_{ST}\cdot \left( \hat{\bf S}_{1} \times \hat{\bf S}_{2}\right).
\label{equ:H}    
\end{equation}
Here, the first term accounts for the Zeeman energy with the $B$-field applied perpendicular to the surface in $z$-direction. 
The second and third terms account for the interaction  between the two intramolecular spins $\hat{\bf S}_{i}=\left(\frac12 \hat \sigma_{x}^{i},\frac12 \hat \sigma_{y}^{i}, \frac12 \hat \sigma_{z}^{i}\right)$ ($\sigma_{x,y,z}$ are the standard Pauli matrices) by an isotropic Heisenberg coupling term $J_{ST}$ and the non-collinear Dzyaloshinskii-Moriya (DM)\cite{dzialoshinskii1957thermodynamic,moriya1960anisotropic} interaction vector $\vec{D}_{ST}$.
We find an excellent agreement between our measured data and simulations which employs equation \ref{equ:H} and a perturbative tunneling model\cite{ternes2017SpinSTM} using a Heisenberg interaction strength of  $J_{ST}=21.6\pm0.5$~meV and a DM interaction vector $\vec{D}_{ST}$ which lies in the surface plane and has a strength of $|\vec{D}_{ST}|=(0.45\pm 0.1)\times J_{ST}$ (Fig.~3d, f).
% 5T: 21.53 meV; 8T: 21.58 meV; 13T: 21.76 meV.
The apparent visibility of only two transitions originates from an asymmetric shift of the triplet state energies so that even at high $B$-fields two of them can not be separated and overlay in the observed $dI/dV$ spectra (Fig.~3h, i). % (Fig.~3g--i and SOM). 
This also clearly exclude that $\vec{D}_{ST}$ has a significant out-of-plane component. 
%To minimze the energy originating from the DM interaction term in equation \ref{equ:H} the two spin vectors are orthogonal to $\vec{D}$ and to each other. 

In contrast to the intermolecular interaction found in layers of CoPc\cite{chen2008PRL_superexchange}, here the main interaction between both spins on the CoPc$_{\rm v}$ molecule is mediated by intramolecular superexchange and varies only slightly ($\pm2.5$\%) with adsorption position on the charge-density-wave modulated 2H-NbSe$_2$ surface (see SOM).
However, the significant DM coupling can not originate from within the flat molecule.
Presumably it is due to interactions between the magnetic moments in the CoPc and the  Nb $d$-orbitals of the 2H-NbSe$_2$\cite{rossnagel2001_CDW_conduction_band_NbSe2_calculation,straub1999_CDW_chargeconduction_band_NbSe2_calculation_ARPES}
%\cite{silva2016Band_structure_NbSe2}
resulting in an in-plane DM vector (Fig.~4a)\cite{khajetoorians2016DMI_Fe_Pt(111)}, in agreement with the experimental data.

To study the excitation of CoPc$_{\rm v}$ in greater detail we take spectra on a grid of %$38 \times 38$ 
points %over an area of $2.5\times 2.5$~nm$^2$ size 
covering one CoPc$_{\rm v}$ molecule. 
At every point we determine $J_{ST}$ assuming a constant ratio $|\vec{D}_{ST}|/J_{ST}=0.45$ and the  intensity of the inelastic conductance relative to the total conductance, $A=\sigma_{\rm inel.}/(\sigma_{\rm el.}+\sigma_{\rm inel.})$ (Fig.~4b,c). 
The $J_{ST}$ map clearly reflects the 4-fold symmetry of the bare molecule (Fig.~4b). 
The observed small variations of $J_{ST}$ with tip position are due to attractive mechanical forces exerted by the tip which bend the molecule and changes thereby the intramolecular magnetic coupling (see SOM).

In stark contrast to the $J_{ST}$ map, the $A$-map shows clear mirror symmetry along the $\sigma_v$ axes of molecule and surface, and a strong variation over the molecule with $A$ ranging from.  $\approx0.5-0.9$.  
This map describes the spatial distribution of the spin excitation intensity, which is correlated to the relative local density of states of the orbitals containing the unpaired  spins\cite{mishra2020_Singlet_Triplet_transition_nanographene}.
Surprisingly, we detect large $A$ not only on the central Co$^{2+}$ ion but also on the phthalocyanine ring as two, c-shaped lobes symmetrically around the mirror plane, clearly marking this direction as the one in which $\vec{D}_{ST}$ lies. 
While part of the detailed sub-structure also depends on the tip apex, we note the faint lines of increased $A$ which link the central Co$^{2+}$ ion via the N-atoms to the benzene rings of the molecule.

To conclude, we have revealed the key role of symmetries between substrate and adsorbate for the spin state and the intramolecular interactions of CoPc molecules on 2H-NbSe$_2$. 
While CoPc$_{\rm d}$ has an unpaired electron in the $d_{z^{2}}$-orbital of the central Co$^{2+}$ ion\cite{kezilebieke2018coupledCoPcNbSe2} which couples to the sample leading either to YSR states or to Kondo screening, the two electron spins in CoPc$_{\rm v}$ couple antiferromagnetically. 
The reduced symmetry and the strong spin-orbit coupling of the 2H-NbSe$_2$ surface induce significant non-collinear DM coupling in CoPc$_{\rm v}$ which lead to an unbalanced field splitting of the singlet-triplet excitation.

%\textcolor{red}{ %I found many misconceptions here. To be physically clear, delocalized state of matter here only exist for the electrons occupying the conduction energy bands of the substrate. In its most simplest form they could be Bloch electrons expanding over the whole lattice and being called as itinerant or delocalized electrons. Any electrons occupying atomic or molecular orbital on the adsorbate is by definition local, regardless of being $\pi$-orbital or $d_{z^{2}}$-orbital, otherwise you could not be able to spatially resolve it via a local probe like STM tip. Together with DMI physics, this is a central point in rationalizing the singlet-triplet excitation with an unbalanced magnetic field splitting in this study and negligence in accurately addressing this issue would undermine the experimental results. }

Our work demonstrates that the spin state of adsorbed molecules can strongly depend on subtle variations of the twist angle with  respect to the substrate which opens a new path for controlling and engineering more complex spin structures. 
%Long in-plane coherence \textcolor{blue}{(of the substrate)} and 
Additionally, the substrate mediated non-collinear interaction in metal-organic molecules is a promising platform for exploiting phenomena such as one-dimensional spin spirals\cite{menzel2012Spin_spiral_Fe_Chain} or topological superconductivity.
%Since 2H-NbSe$_2$ is a layered structure, YSR states have a long in-plane range than those of three-dimensional superconductors.
%It extends to design building more complex spins structure by molecular manipulation.
%Our results provides a platform for probing intramolecular spin interaction, and the possibility to study the DMI in a zero dimensional molecule of two simple $S=1/2$ spins, which is mostly studied in 2D system, for example, magnetic skymions\cite{romming2013skymions}, or 1D system, for example, one-dimensional spin spiral\cite{menzel2012Spin_spiral_Fe_Chain}.
%However, further theoretical and experimental research are required to understand how the symmetry breaking induces DMI in the intra-molecular magnetic coupling.
%\textcolor{red}{(S: My humble view; let's less oversell the results. It is already advertised twice in the abstract and introduction (from topological superconductivity to skyrmions and Majorana physics, WOW! It is just a singlet-triplet excitation interplaying with DMI!).}

\begin{methods}
\textbf{Experimental procedure.}
The 2H-NbSe$_2$ single crystal was cleaved by attaching an adhesive Kapton polyimide tape to the crystal surface and pulling it off at a base pressure of $p\leq 10^{-8}$~mbar. 
CoPc molecules were then deposited from a Knudsen cell evaporator held at 410\textcelsius\ onto the freshly cleaved 2H-NbSe$_2$ at room temperature and $p\leq 10^{-9}$~mbar.
The SPM experiments were performed using a home-built combined scanning tunneling and atomic force microscope operating in ultrahigh vacuum ($p\leq 10^{-10}$~mbar), at fields perpendicular to the sample surface of up to 14~T, and at a base temperature of $1.2$~K. 
The differential conductance ($dI/dV$) spectra were detected by modulating the bias voltage $V$ with a sinusoidal of $0.05-0.2$~mV amplitude and $617$~Hz frequency utilizing a lock-in amplifier. 
We functionalized the bare Pt tip by indenting it into a Pb surface by several hundreds of nm repeatedly until it showed a bulk-like superconducting gap. 
The tip is mounted on a quartz tuning fork with a resonance frequency of $f_0= 29,067$~Hz, a stiffness of $k=1800$~N/m, and a $Q$-factor of $\approx 60,000$.
Tuning fork oscillation amplitudes of $50$~pm were used to measure the forces acting between tip and sample by detecting the frequency shift $df$ of the tuning fork.

%\noindent\textbf{Ab-initio calculations.}
%The simulations were based on density functional theory (DFT) on the basis of the plane-wave implementation in the Quantum Espresso~\cite{QE2009,QE2017} package. Projector augmented pseudopotentials were used [ref?] while the plane waves energy cutoff was set to 400 Ry. The exchange-correlations effects were treated utilizing the PBE~\cite{}[ref] generalized gradient functional. For the free CoPC molecule, a unit cell of 36$\times$36 angstrom was set, and the Brillouin zone sampled with a single k-point. For the geometrical relaxations, a Gaussian smearing of 0.02 Ry and a convergence threshold of 1e-3 a.u. was established. For the molecule on the surface, an 8x8 monolayer NbSe2 supercell was constructed with vdW interactions taken into account in the atomic relaxation procedure [ref to Grimme’s DFT-D3 method]. Electronic correlation effects were accounted for utilizing a $U$ of ... eV~[ref:DFT+U method].
%Here is a description of a specific method used.  Note that the subsection heading ends with a full stop (period) and that the command is \verb|\subsection{}| not \verb|\subsection*{}|.

\end{methods}

%% Put the bibliography here, most people will use BiBTeX in
%% which case the environment below should be replaced with
%% the \bibliography{} command.

\subsection{References}
%
%\bibliography{bibliography.bib}

\begin{thebibliography}{10}
\expandafter\ifx\csname url\endcsname\relax
  \def\url#1{\texttt{#1}}\fi
\expandafter\ifx\csname urlprefix\endcsname\relax\def\urlprefix{URL }\fi
\providecommand{\bibinfo}[2]{#2}
\providecommand{\eprint}[2][]{\url{#2}}

\bibitem{cao2018unconventional}
\bibinfo{author}{Cao, Y.} \emph{et~al.}
\newblock \bibinfo{title}{Unconventional superconductivity in magic-angle
  graphene superlattices}.
\newblock \emph{\bibinfo{journal}{Nature}} \textbf{\bibinfo{volume}{556}},
  \bibinfo{pages}{43--50} (\bibinfo{year}{2018}).

\bibitem{cao2018correlated}
\bibinfo{author}{Cao, Y.} \emph{et~al.}
\newblock \bibinfo{title}{Correlated insulator behaviour at half-filling in
  magic-angle graphene superlattices}.
\newblock \emph{\bibinfo{journal}{Nature}} \textbf{\bibinfo{volume}{556}},
  \bibinfo{pages}{80} (\bibinfo{year}{2018}).

\bibitem{sharpe2019ferromagnetism_in_twisted_bilayer_graphene}
\bibinfo{author}{Sharpe, A.~L.} \emph{et~al.}
\newblock \bibinfo{title}{Emergent ferromagnetism near three-quarters filling
  in twisted bilayer graphene}.
\newblock \emph{\bibinfo{journal}{Science}} \textbf{\bibinfo{volume}{365}},
  \bibinfo{pages}{605--608} (\bibinfo{year}{2019}).

\bibitem{chen2019Mott_insulator_trilayer_graphene}
\bibinfo{author}{Chen, G.} \emph{et~al.}
\newblock \bibinfo{title}{Evidence of a gate-tunable {Mott} insulator in a
  trilayer graphene moir{\'e} superlattice}.
\newblock \emph{\bibinfo{journal}{Nat. Phys.}} \textbf{\bibinfo{volume}{15}},
  \bibinfo{pages}{237--241} (\bibinfo{year}{2019}).

\bibitem{
chen2019gate_tunable_superconductivity_in_trilayer_graphene_hBN_moire_superlatti
ces}
\bibinfo{author}{Chen, G.} \emph{et~al.}
\newblock \bibinfo{title}{Signatures of tunable superconductivity in a trilayer
  graphene moir{\'e} superlattice}.
\newblock \emph{\bibinfo{journal}{Nature}} \textbf{\bibinfo{volume}{572}},
  \bibinfo{pages}{215--219} (\bibinfo{year}{2019}).

\bibitem{mugarza2012DFT_MPC_gasphase_onAg(100)}
\bibinfo{author}{Mugarza, A.} \emph{et~al.}
\newblock \bibinfo{title}{Electronic and magnetic properties of molecule-metal
  interfaces: Transition-metal phthalocyanines adsorbed on {Ag}(100)}.
\newblock \emph{\bibinfo{journal}{Phys. Rev. B}} \textbf{\bibinfo{volume}{85}},
  \bibinfo{pages}{155437} (\bibinfo{year}{2012}).

\bibitem{xi2016NbSe2isingSC}
\bibinfo{author}{Xi, X.} \emph{et~al.}
\newblock \bibinfo{title}{Ising pairing in superconducting $\text{NbSe}_{2}$
  atomic layers}.
\newblock \emph{\bibinfo{journal}{Nat. Phys.}} \textbf{\bibinfo{volume}{12}},
  \bibinfo{pages}{139--143} (\bibinfo{year}{2016}).

\bibitem{dzialoshinskii1957thermodynamic}
\bibinfo{author}{Dzialoshinskii, I.}
\newblock \bibinfo{title}{Thermodynamic theory of weak ferromagnetism in
  antiferromagnetic substances}.
\newblock \emph{\bibinfo{journal}{Sov. Phys. JETP}}
  \textbf{\bibinfo{volume}{5}}, \bibinfo{pages}{1259--1272}
  (\bibinfo{year}{1957}).

\bibitem{moriya1960anisotropic}
\bibinfo{author}{Moriya, T.}
\newblock \bibinfo{title}{Anisotropic superexchange interaction and weak
  ferromagnetism}.
\newblock \emph{\bibinfo{journal}{Phys. Rev.}} \textbf{\bibinfo{volume}{120}},
  \bibinfo{pages}{91} (\bibinfo{year}{1960}).

\bibitem{cotton2003chemical}
\bibinfo{author}{Cotton, F.~A.}
\newblock \emph{\bibinfo{title}{Chemical applications of group theory}}
  (\bibinfo{publisher}{John Wiley \& Sons}, \bibinfo{year}{2003}).

\bibitem{menard2015coherent_length_YSR_Co_NbSe2}
\bibinfo{author}{M{\'e}nard, G.~C.} \emph{et~al.}
\newblock \bibinfo{title}{Coherent long-range magnetic bound states in a
  superconductor}.
\newblock \emph{\bibinfo{journal}{Nat. Phys.}} \textbf{\bibinfo{volume}{11}},
  \bibinfo{pages}{1013--1016} (\bibinfo{year}{2015}).

\bibitem{bode2007chiral}
\bibinfo{author}{Bode, M.} \emph{et~al.}
\newblock \bibinfo{title}{Chiral magnetic order at surfaces driven by inversion
  asymmetry}.
\newblock \emph{\bibinfo{journal}{Nature}} \textbf{\bibinfo{volume}{447}},
  \bibinfo{pages}{190--193} (\bibinfo{year}{2007}).

\bibitem{kezilebieke2018coupledCoPcNbSe2}
\bibinfo{author}{Kezilebieke, S.}, \bibinfo{author}{Dvorak, M.},
  \bibinfo{author}{Ojanen, T.} \& \bibinfo{author}{Liljeroth, P.}
\newblock \bibinfo{title}{Coupled $\text{Yu-Shiba-Rusinov}$ states in molecular
  dimers on $\text{NbSe}_{2}$}.
\newblock \emph{\bibinfo{journal}{Nano Lett.}} \textbf{\bibinfo{volume}{18}},
  \bibinfo{pages}{2311--2315} (\bibinfo{year}{2018}).

\bibitem{yu1965_Y_YSR}
\bibinfo{author}{Yu, L.}
\newblock \bibinfo{title}{Bound state in superconductors with paramagnetic
  impurities}.
\newblock \emph{\bibinfo{journal}{Acta Phys. Sin.}}
  \textbf{\bibinfo{volume}{21}}, \bibinfo{pages}{75--91}
  (\bibinfo{year}{1965}).

\bibitem{shiba1968_S_YSR}
\bibinfo{author}{Shiba, H.}
\newblock \bibinfo{title}{Classical spins in superconductors}.
\newblock \emph{\bibinfo{journal}{Prog. Theor. Exp. Phys.}}
  \textbf{\bibinfo{volume}{40}}, \bibinfo{pages}{435--451}
  (\bibinfo{year}{1968}).

\bibitem{rusinov1969_R_YSR}
\bibinfo{author}{Rusinov, A.}
\newblock \bibinfo{title}{Theory of gapless superconductivity in alloys
  containing paramagnetic impurities}.
\newblock \emph{\bibinfo{journal}{JETP Lett.}} \textbf{\bibinfo{volume}{29}},
  \bibinfo{pages}{1101--1106} (\bibinfo{year}{1969}).

\bibitem{heinrich2018YSRreview}
\bibinfo{author}{Heinrich, B.~W.}, \bibinfo{author}{Pascual, J.~I.} \&
  \bibinfo{author}{Franke, K.~J.}
\newblock \bibinfo{title}{Single magnetic adsorbates on s-wave
  superconductors}.
\newblock \emph{\bibinfo{journal}{Prog. Surf. Sci.}}
  \textbf{\bibinfo{volume}{93}}, \bibinfo{pages}{1--19} (\bibinfo{year}{2018}).

\bibitem{salkola2002YSR_fit}
\bibinfo{author}{Salkola, M.}, \bibinfo{author}{Balatsky, A.} \&
  \bibinfo{author}{Schrieffer, J.}
\newblock \bibinfo{title}{Spectral properties of quasiparticle excitations
  induced by magnetic moments in superconductors}.
\newblock \emph{\bibinfo{journal}{Phys. Rev. B}} \textbf{\bibinfo{volume}{55}},
  \bibinfo{pages}{12648} (\bibinfo{year}{1997}).

\bibitem{Zhang13_split_Kondo}
\bibinfo{author}{Zhang, Y.} \emph{et~al.}
\newblock \bibinfo{title}{Temperature and magnetic field dependence of a
  \text{Kondo} system in the weak coupling regime}.
\newblock \emph{\bibinfo{journal}{Nat. Commun.}} \textbf{\bibinfo{volume}{4}},
  \bibinfo{pages}{2110} (\bibinfo{year}{2013}).

\bibitem{kondo1968effect}
\bibinfo{author}{Kondo, J.}
\newblock \bibinfo{title}{Effect of ordinary scattering on exchange scattering
  from magnetic impurity in metals}.
\newblock \emph{\bibinfo{journal}{Phys. Rev.}} \textbf{\bibinfo{volume}{169}},
  \bibinfo{pages}{437} (\bibinfo{year}{1968}).

\bibitem{ternes2017SpinSTM}
\bibinfo{author}{Ternes, M.}
\newblock \bibinfo{title}{Probing magnetic excitations and correlations in
  single and coupled spin systems with scanning tunneling spectroscopy}.
\newblock \emph{\bibinfo{journal}{Prog. Surf. Sci.}}
  \textbf{\bibinfo{volume}{92}}, \bibinfo{pages}{83--115}
  (\bibinfo{year}{2017}).

\bibitem{vzitko2009splitting_Kondo}
\bibinfo{author}{{\v{Z}}itko, R.}, \bibinfo{author}{Peters, R.} \&
  \bibinfo{author}{Pruschke, T.}
\newblock \bibinfo{title}{Splitting of the \text{Kondo} resonance in
  anisotropic magnetic impurities on surfaces}.
\newblock \emph{\bibinfo{journal}{New J. Phys.}} \textbf{\bibinfo{volume}{11}},
  \bibinfo{pages}{053003} (\bibinfo{year}{2009}).

\bibitem{franke2011competition_YSR_Kondo}
\bibinfo{author}{Franke, K.}, \bibinfo{author}{Schulze, G.} \&
  \bibinfo{author}{Pascual, J.}
\newblock \bibinfo{title}{Competition of superconducting phenomena and
  \text{Kondo} screening at the nanoscale}.
\newblock \emph{\bibinfo{journal}{Science}} \textbf{\bibinfo{volume}{332}},
  \bibinfo{pages}{940--944} (\bibinfo{year}{2011}).

\bibitem{heinrich2013Spin_excitation_SC_tip}
\bibinfo{author}{Heinrich, B.}, \bibinfo{author}{Braun, L.},
  \bibinfo{author}{Pascual, J.} \& \bibinfo{author}{Franke, K.}
\newblock \bibinfo{title}{Protection of excited spin states by a
  superconducting energy gap}.
\newblock \emph{\bibinfo{journal}{Nat. Phys.}} \textbf{\bibinfo{volume}{9}},
  \bibinfo{pages}{765--768} (\bibinfo{year}{2013}).

\bibitem{chiang2014IETS_COtip}
\bibinfo{author}{Chiang, C.-l.}, \bibinfo{author}{Xu, C.},
  \bibinfo{author}{Han, Z.} \& \bibinfo{author}{Ho, W.}
\newblock \bibinfo{title}{Real-space imaging of molecular structure and
  chemical bonding by single-molecule inelastic tunneling probe}.
\newblock \emph{\bibinfo{journal}{Science}} \textbf{\bibinfo{volume}{344}},
  \bibinfo{pages}{885--888} (\bibinfo{year}{2014}).

\bibitem{chen2008PRL_superexchange}
\bibinfo{author}{Chen, X.} \emph{et~al.}
\newblock \bibinfo{title}{Probing superexchange interaction in molecular
  magnets by spin-flip spectroscopy and microscopy}.
\newblock \emph{\bibinfo{journal}{Phys. Rev. Lett.}}
  \textbf{\bibinfo{volume}{101}}, \bibinfo{pages}{197208}
  (\bibinfo{year}{2008}).

\bibitem{rossnagel2001_CDW_conduction_band_NbSe2_calculation}
\bibinfo{author}{Rossnagel, K.} \emph{et~al.}
\newblock \bibinfo{title}{Fermi surface of \text{2H}-$\text{NbSe}_{2}$ and its
  implications on the charge-density-wave mechanism}.
\newblock \emph{\bibinfo{journal}{Phys. Rev. B}} \textbf{\bibinfo{volume}{64}},
  \bibinfo{pages}{235119} (\bibinfo{year}{2001}).

\bibitem{straub1999_CDW_chargeconduction_band_NbSe2_calculation_ARPES}
\bibinfo{author}{Straub, T.} \emph{et~al.}
\newblock \bibinfo{title}{Charge-density-wave mechanism in
  \text{2H}-$\text{NbSe}_{2}$: Photoemission results}.
\newblock \emph{\bibinfo{journal}{Phys. Rev. Lett.}}
  \textbf{\bibinfo{volume}{82}}, \bibinfo{pages}{4504} (\bibinfo{year}{1999}).

\bibitem{khajetoorians2016DMI_Fe_Pt(111)}
\bibinfo{author}{Khajetoorians, A.} \emph{et~al.}
\newblock \bibinfo{title}{Tailoring the chiral magnetic interaction between two
  individual atoms}.
\newblock \emph{\bibinfo{journal}{Nat. Commun.}} \textbf{\bibinfo{volume}{7}},
  \bibinfo{pages}{1--8} (\bibinfo{year}{2016}).

\bibitem{mishra2020_Singlet_Triplet_transition_nanographene}
\bibinfo{author}{Mishra, S.} \emph{et~al.}
\newblock \bibinfo{title}{Topological frustration induces unconventional
  magnetism in a nanographene}.
\newblock \emph{\bibinfo{journal}{Nat. Nanotechnol.}}
  \textbf{\bibinfo{volume}{15}}, \bibinfo{pages}{22--28}
  (\bibinfo{year}{2020}).

\bibitem{menzel2012Spin_spiral_Fe_Chain}
\bibinfo{author}{Menzel, M.} \emph{et~al.}
\newblock \bibinfo{title}{Information transfer by vector spin chirality in
  finite magnetic chains}.
\newblock \emph{\bibinfo{journal}{Phys. Rev. Lett.}}
  \textbf{\bibinfo{volume}{108}}, \bibinfo{pages}{197204}
  (\bibinfo{year}{2012}).

\end{thebibliography}

\begin{addendum}
 \item [Acknowledgements] We thank Christian Ast, Haonan Huang,  Shawulienu Kezilebieke, Peter Liljeroth, Samir Lounis, Ana Montero, and Lihui Zhou for fruitful discussions. 
 %Filipe Guimaraes, Ruslan Temirov
 %Haonan Huang, Christian Ast: fit the YSR spectra. Haonan helps with the section 2, SI
 % Lihui Zhou: help improve the energy resolution, the NbSe2 is from him, share the result which is not on the paper: magnetic atom doesn't show YSR states but shows an excitation.
 % Abhishek Grewal, Klaus Kuhnke: molecular orbitals and mass spec;
 % Sujoy Karan: Charge state of molecule;
 % Xu Wu: Molecular spin; 
 Y.\,W.\ acknowledges support from the Alexander von Humboldt Foundation and M.\,T.\ by the Heisenberg Program (Grant No.\ TE 833/2-1) of the German Research Foundation.
 \item[Author contributions] M.\,T.\ conceived the experiment. Y.\,W.\ and S.\,A.\ performed the SPM measurements. Y.\,W.\ and M.\,T.\  analysed and fitted the data. M.\,T.\ and K.\,K.\ supervised the project. All authors discussed the results and contributed to the manuscript.
 \item[Competing financial interests] The authors declare no competing financial interests.
 \item[Data availability] The relevant spectroscopic data sets used in this publication are available from the authors.
\end{addendum}

\clearpage

\begin{figure}
\includegraphics[width=0.99\textwidth]{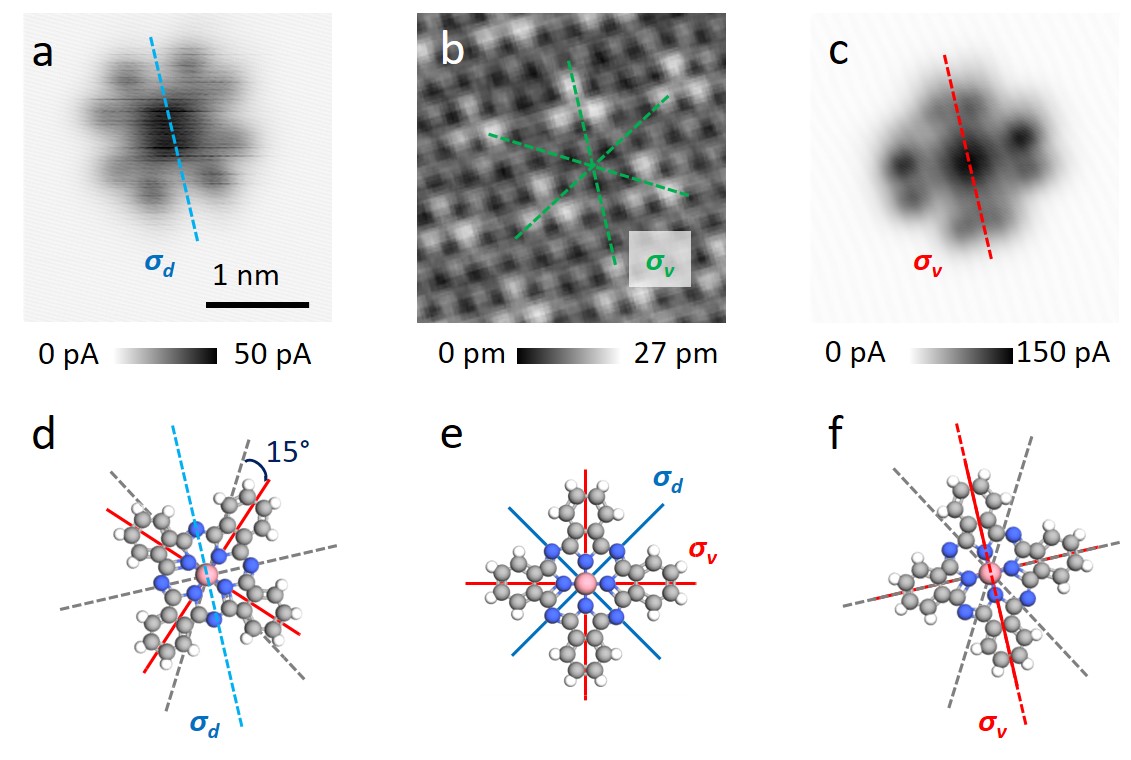}
\caption{\textbf{Absorption symmetry and spectral fingerprint of CoPc molecules on 2H-NbSe$_2$.}
\textbf{a, c}, Constant height SPM images of two CoPc molecules ($V=400$~mV) adsorbed in different orientations. 
\textbf{b}, Constant current image of the 2H-NbSe$_2$ surface showing the $3\times3$ charge density wave superstructure ($V=-10$~mV, $I=1$ nA). 
Colored dashed lines in \textbf{a--c} mark the different mirror planes of the CoPc molecules and the 2H-NbSe$_2$ surface. 
\textbf{d, f}, Absorption models of CoPc molecules on 2H-NbSe$_2$. 
While the molecule in (\textbf{d}) is rotated by 15$^\circ$ with respect to one of the three principal axes of the substrate (grey dashed lines), the molecule in (\textbf{f}) is aligned. \textbf{e}, Model of the CoPc molecule with its vertical $\sigma_v$ (red lines) and diagonal $\sigma_d$  (blue lines) mirror plane symmetries.
%\textbf{g}, Schematic illustration of CoPc$_{\rm KD}$ containing a Kramer's doublet (KD) coupled to the Cooper pairs of the substrate.
%\textbf{h}, Schematic illustration of the intramolecular coupling between the Co$^{2+}$ central ion and the ligands in CoPc$_{\rm ST}$ leading to an $S=0$ singlet to $S=1$ triplet (ST) transition.
}
\end{figure}

\begin{figure}
\includegraphics[width=\textwidth]{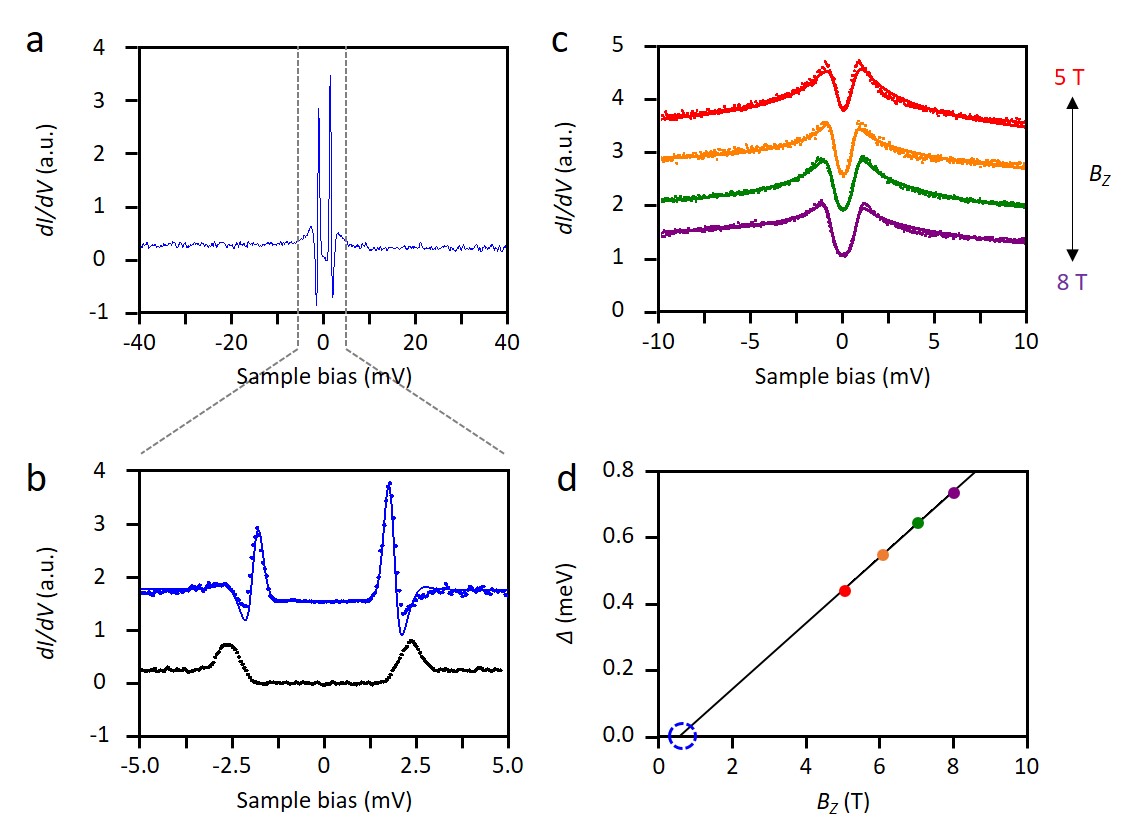}
\caption{\textbf{Spectral features of CoPc$_{\rm d}$}. \textbf{a, b}, Differential conductance $dI/dV$ measured on the bare NbSe$_2$ sample (black dots) and the center of a CoPc$_{\rm d}$ molecule (blue dots) by using a SC tip ($V=-40$~mV, $I=40$~pA in (\textbf{a}); $V=-5$~mV, $I=50$~pA in (\textbf{b})). 
Full line in (\textbf{b}) is a least-square fit to a scattering model in which the magnetic impurity is treated classically. \textbf{c}, $dI/dV$ spectra measured on CoPc$_{\rm d}$ at magnetic fields large enough to suppress SC (dotted lines, $V=-10$~mV, $I=100$~pA) and least-square fits using a perturbative scattering model (full lines).
Curves are vertically offset for clarity. 
\textbf{d}, Extracted splitting of the peaks in (\textbf{c}) and linear 
regression (full line). The dashed circle marks the crossing of the regression 
with the abscissa.}
\end{figure}

\begin{figure}
    \centering
\includegraphics[width=0.7\textwidth]{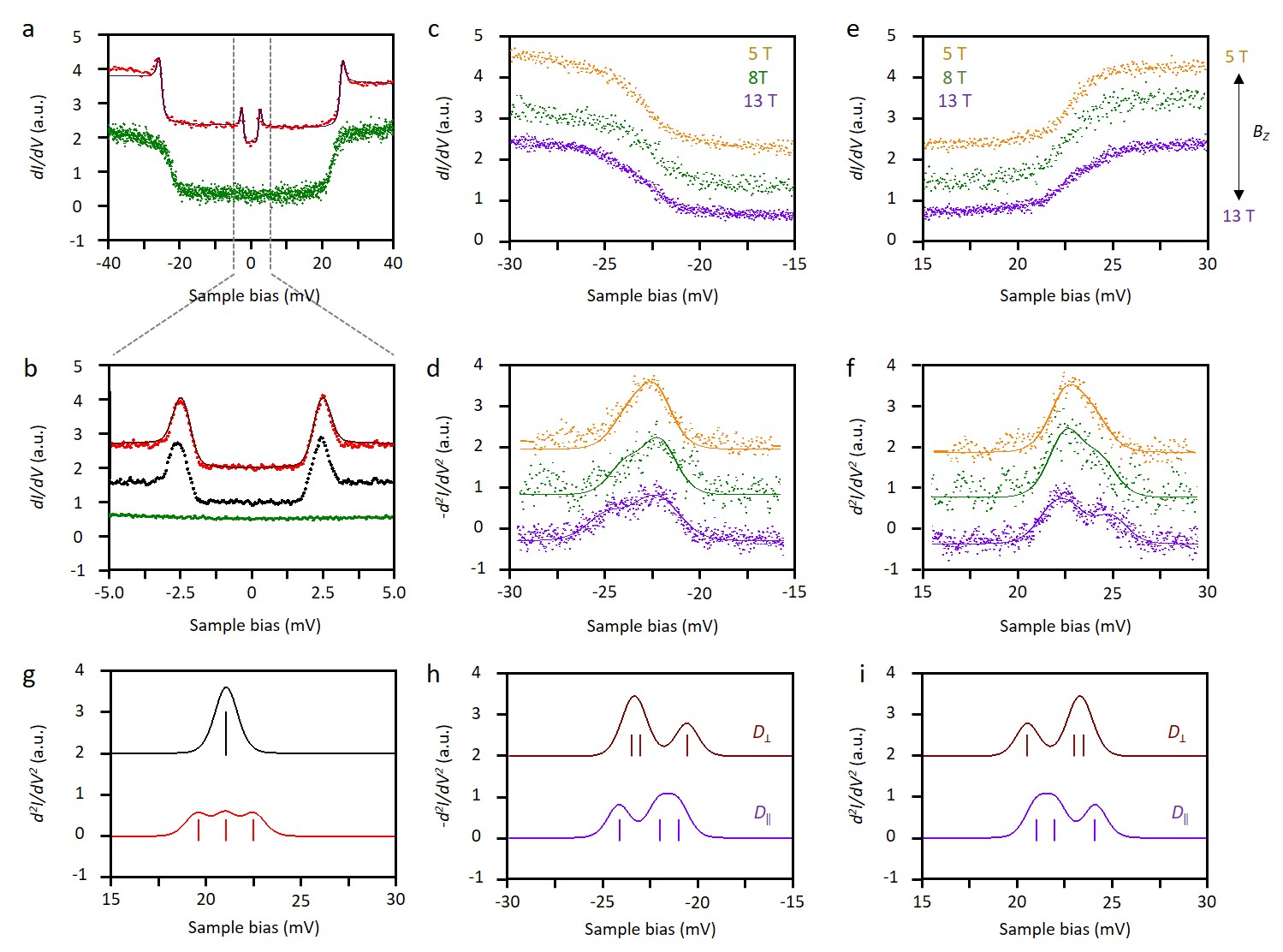}
\caption{\textbf{The singlet -- triplet transition in CoPc$_{\rm v}$}. \textbf{a}, $dI/dV$ spectra measured on CoPc$_{\rm v}$ at $B=0$ (red dots) and at $B=8$~T where SC is quenched (green dots). 
The full line at $B=0$ is a least-square fit to a model which accounts for the SC gaps in tip and sample and the spin excitation. 
\textbf{b}, Detail of the curves in (\textbf{a}) (red and green dots) and spectrum measured on the bare NbSe$_2$ surface (black dots), showing that neither YSR states nor a Kondo peak can be detected on the molecule ($V=-5$ mV, $I=50$ pA). 
The dark red line is a fit to a SC--SC tunneling model.
\textbf{c--f}, $dI/dV$ and numerically derived $d^{2}I/dV^{2}$ spectra measured on CoPc$_{\rm v}$ at $B=5,8$ and $13$~T, respectively.
Full lines in (\textbf{d,f}) are least-square fit to a perturbative transport model.
The spectra reveal an asymmetric splitting of the inelastic excitation at $\approx \pm23$~mV in field.
\textbf{g}, Expected splitting of a triplet excitation at $B=13$~T (red line) if only Heisenberg exchange interaction between both spins is taken into account. 
\textbf{h, i}, Accounting for an additional non-collinear Dzyaloshinskii–Moriya (DM) interaction rationalized the observation if the DM vector lies in the surface plane (D$_{||}$).
A DM vector pointing out of surface (D$_\bot$) would reverse the intensity order.  Curves in all panels are vertically offset for clarity.}
\end{figure}

\begin{figure}
    \centering
\includegraphics[width=\textwidth]{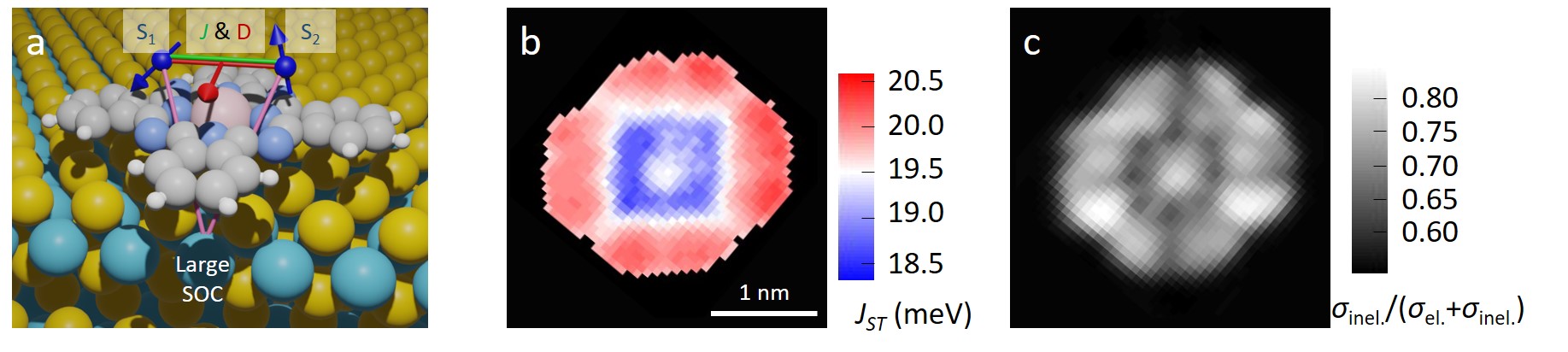}
\caption{\textbf{Spin excitation map of CoPc$_{\rm v}$}. 
\textbf{a}, Schematic ball model of the CoPc$_{\rm v}$
and its interactions. 
Grey, white, light blue, and pink spheres correspond to C, H, N, and Co atoms, on the molecule, respectively. 
Yellow and turquoise spheres correspond to Se and Nb atoms of the surface. 
Blue arrows indicate the two spins and the red arrow the DM vector. 
\textbf{b,c}, Maps of 45$\times$45 points covering an area of 3 $\times$ 3 nm$^2$ on which  $dI/dV$ spectra where taken and \textbf{b} the interaction strength $J_{ST}$ and \textbf{c} the spin excitation intensity $A=\sigma_{\rm inel.}/(\sigma_{\rm inel.}+\sigma_{\rm el.})$ where extracted ($V=-50$~mV, $I=500$~pA).
While $J_{ST}$ map shows mainly fourfold symmetry, $A$-map clearly reveals the mirror plane which cuts approximately vertical through the image.
}

\end{figure}

\end{document}